\documentclass[12pt,preprint]{aastex}
\begin{document}
\title{GRS 1915+105 as a Galactic Analog of a Fanaroff-Riley II  Quasar}
\author{Brian Punsly\altaffilmark{1}} \and\author{J\'{e}r$\hat{\mathrm{o}}$me Rodriguez\altaffilmark{2}}
\altaffiltext{1}{1415 Granvia Altamira, Palos Verdes Estates CA, USA
90274 and ICRANet, Piazza della Repubblica 10 Pescara 65100, Italy,
brian.punsly1@verizon.net or brian.punsly@comdev-usa.com}
\altaffiltext{2}{Laboratoire AIM, CEA/DSM-CNRS-Universit\'{e} Paris
Diderot, IRFU SAp, F-91191 Gif-sur-Yvette, France.}
\begin{abstract}
We study the long term time averaged kinetic luminosity,
$\overline{Q}$, of the major flares of the Galactic microquasar GRS
1915+105 and the relationship to the intrinsic X-ray (bolometric)
luminosity, $L_{\mathrm{bol}}$, and scale it to that of a complete
sample of SDSS/FIRST FR II quasars. If the scale invariance
hypothesis for black holes (BHs) holds then we show that the
expected distribution in the $\overline{Q}$ - $L_{\mathrm{bol}}$
scatter plane of GRS 1915+105 is consistent with FR II quasars for
distances D = 10.7 - 11 kpc. We compare the specific values of
kinetic luminosity and $L_{\mathrm{bol}}$ during flares of GRS
1915+105 to that predicted by several 3-D MHD simulations of BH
accretion flows with relativistic ejections. If FR II quasars are a
scaled up version of GRS 1915+105, the data are consistent with
numerical models when they contain an ergospheric disk jet and the
BH spin is $a/M= 0.99$ or $a/M=0.998$ (we estimate $a/M>0.984$). In the framework of scale invariance of BHs, our results may imply that FR
II quasars also hold rapidly rotating BHs.
\end{abstract}


\section{Introduction}
The black hole candidate GRS~1915+105 is well known for launching
superluminal radio flares out to large distances at a much larger
rate than any other Galactic object
\citep{mir94,fen99,dha00}. The X-ray luminosity of
GRS~1915+105 is one of the highest of any known Galactic black hole
(BH) candidate \citep{don04}. The existence of both relativistic
outflows and high continuum luminosity make it tempting to speculate
that GRS 1915+105 might be a scaled down version of a radio loud
quasar. This is of particular importance because the time scales for
radio evolution are reduced from the AGN (active galactic nuclei)
time scales by many orders of magnitude. Thus, unlike quasars, it is
in principle possible to see the details of the connection between
the putative accretion flow (X-ray luminosity) and the superluminal
jet launching mechanism.
\par In this article, the idea of GRS 1915+105 as a scaled
down FR(Fanaroff-Riley) II quasar is explored. The long term time averaged power of
the relativistic major flares in GRS 1915+105, $\overline{Q}$, and
the luminosity associated with viscous dissipation in the accretion
flow, $L_{\mathrm{bol}}$, are re-scaled in order to compare and
contrast with the distribution of $\overline{Q}$ and
$L_{\mathrm{bol}}$ of a complete sample of FR II quasars. These
efforts are rendered credible by the recent study of the power
required to launch individual major flares and their accretion state
(intrinsic X-ray luminosity or bolometric luminosity,
$L_{\mathrm{bol}}$) just hours and minutes before ejection and
during the brief 1 to 7 hour ejection event (Punsly and Rodrgiuez (2013), PR13
hereafter). The relevant results from PR13 that are required to
perform this re-scaling are indicated in Section 2. In Section 3,
this is compared to FR II quasars. In Section 4, the results of
Sections 2 and 3 are considered in the context of numerical
simulations of 3-D MHD (magnetohydrodynamic) accretion around
spinning BHs.
\section{Estimating the Energy Output from Relativistic Ejecta in GRS 1915+105}
It was determined in \citet{pun12}, P12, that knowledge of the time
evolution of the spectral shape associated with a changing
synchrotron - self absorbed (SSA) opacity, $\tau$, greatly enhances
the accuracy of plasmoid energy estimates (constrains the size). The
frequency and the width of the spectral peak provide two added
pieces of information at each epoch of observation beyond the single
epoch spectral index and flux density of the optically thin high
frequency tail that is traditionally used to estimate the ejected
plasmoid energy. The evolving $\tau$ combined with baryon number
conservation, energy conservation, synchrotron cooling times and
X-ray luminosity were used in P12 to eliminate uncertainty in the
energy estimates. Namely, the proton content is minimal and a near
minimum energy condition, $E_{min}\approx m_{e}c^{2}$, is shown to
occur when the optically thin flux at 2.3 GHz,
$S_{\mathrm{thin}}(2.3)$, is near maximum and $\tau \approx 0.1$. In
PR13 we assume that the detailed modeling of the time evolution of
the flares from P12 can be used as a template for the time evolution
of other plasmoids with less supporting data. This determines a set
of equations for each flare in PR13 that are solved numerically with
4 inputs, the peak $S_{\mathrm{thin}}(2.3)$, the spectral index of
the optically thin emission, D, and the Doppler factor, $\delta$.
Figure 1 is a plot of the plasmoid energy estimated in Table 2 of
PR13 as a function of the estimated peak $S_{\mathrm{thin}}(2.3)$
for the major flares that were listed in Table 1 of PR13. Except when noted, a
fiducial distance to GRS 1915+105 of D =11 kpc is assumed throughout
the manuscript. Unlike Table 2 of PR13, the energy is measured in
the observer's reference frame instead of the plasmoid reference
frame. The power law fit in Figure 1 was made with the method of
weighted least squares with errors in both variables \citep{ree89}.
The corresponding power law fit for each D value is used here to
estimate flare energy from a single input, the peak
$S_{\mathrm{thin}}(2.3)$. The approximately daily 2.3 GHz and 8.3
GHz monitoring with GBI (Green Bank Interferometer) provides a
database from 1996 through 2000 of 1967 days (taking account of gaps
in coverage) for which one can determine $S_{\mathrm{thin}}(2.3)$.
The resulting distributions of major flare energy and number from
1996 through 2000 are plotted in Figure 2.

\par There are three major sources of uncertainty in the long term cumulative energy, $E$, from the flares
\begin{enumerate}
\item The uncertainty in the estimate of the plasmoid energy
\item The minimal $S_{\mathrm{thin}}(2.3)$ that is indicative of a superluminal ejection
\item Possible short (weak) flares missed in the intra-day gaps in coverage
\end{enumerate}

\begin{figure}
\includegraphics[width=160 mm, angle= 0]{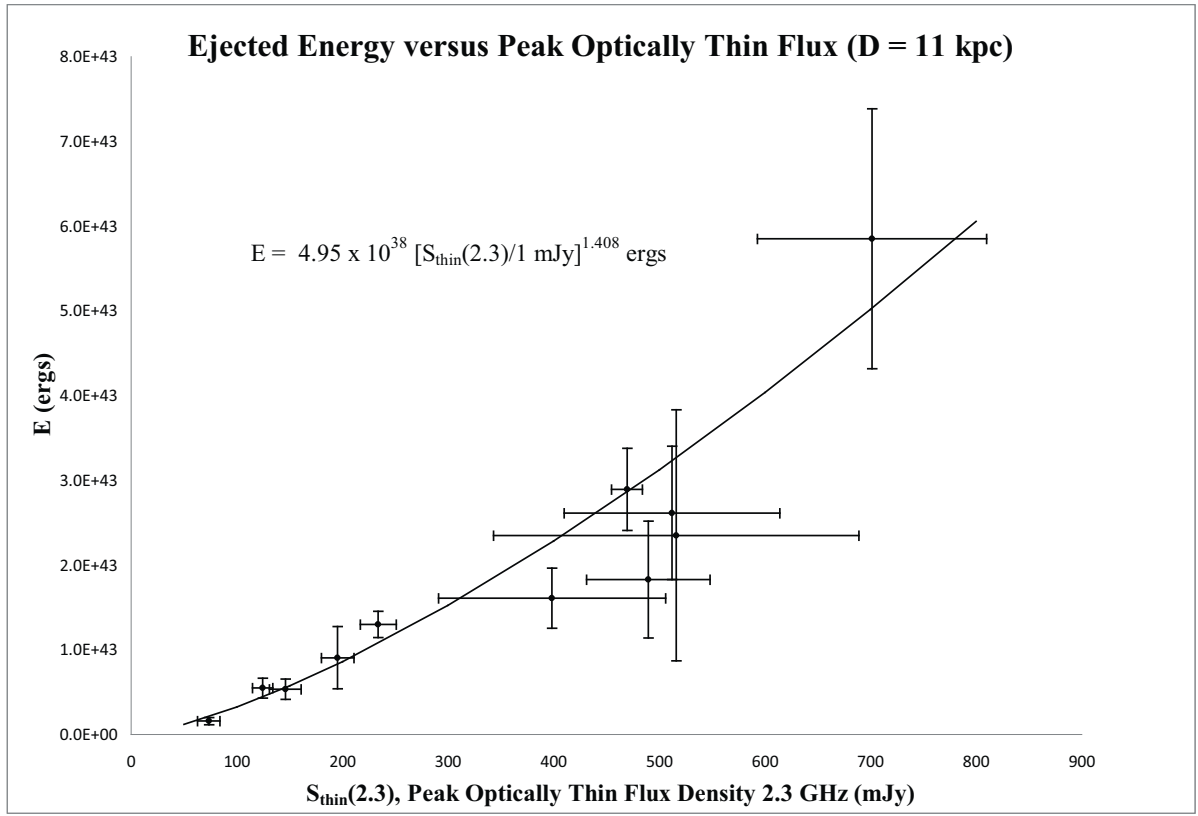}
\caption{This scatter plot of the estimated energy of superluminal
ejections from PR13 evaluated in the observer reference frame versus
the estimated peak $S_{\mathrm{thin}}(2.3)$}
\end{figure}

The stochastic error in the first item is ignored because there are sufficient flares to
drive the propagated random error to approximately 0. We consider every plausible value of
D and the corresponding dependent $\delta$ because of the large systematic uncertainty in $\delta$.
In order to estimate $\delta$ from D, we assume that the kinematic results from \citet{fen99}
are common to the entire time frame from 1997 to 2000 as evidenced
by interferometric observations of multiple flares
\citep{dha00,mil05}. The \emph{intrinsic} spectral luminosity is
$S_{\mathrm{thin}}(2.3)\delta^{-(3+\alpha)}$. As D is varied from
10.5 kpc to near the maximum kinematically allowed value of 11 kpc,
the intrinsic spectral luminosity changes by a factor
$\approx(0.31/0.54)^{3.9}=(1/8.7)$ which equates to a reduction of
the plasmoid energy by a factor of 5 - 6. Alternatively, the
discussions to follow can be phrased in terms of $\delta$ instead of
D. For item 2, we assume that the weakest flares have a peak
$S_{\mathrm{thin}}(2.3)=30\;\mathrm{mJy}$ since this is the smallest
value that can be clearly discerned from a background consisting of
core flux variations and previous fading flares. Extrapolating the distributions in Figure 2 to lower cutoffs shows that the total energy output is
rather insensitive to the low energy cutoff of the flares - most of
the flares are weak, but they carry a small fraction of the total
energy output. We estimate the contribution from weak flares in the
intra-day gaps and the uncertainty from points 2 and 3 to compute
$\overline{Q}$,

\begin{eqnarray}
&&\overline{Q} = \frac{\mathrm{cumulative\;energy\;of\;major\;ejections}}{\mathrm{1967 \;days}} \nonumber \\
&&=\frac{1.78 \pm 0.21 \times 10^{45}\mathrm{ergs}} {\mathrm{1967
\;days}} = 1.04 \pm 0.13 \times 10^{37}\mathrm{ergs/s}\,.
\end{eqnarray}

\begin{figure}
\includegraphics[width=160 mm, angle= 0]{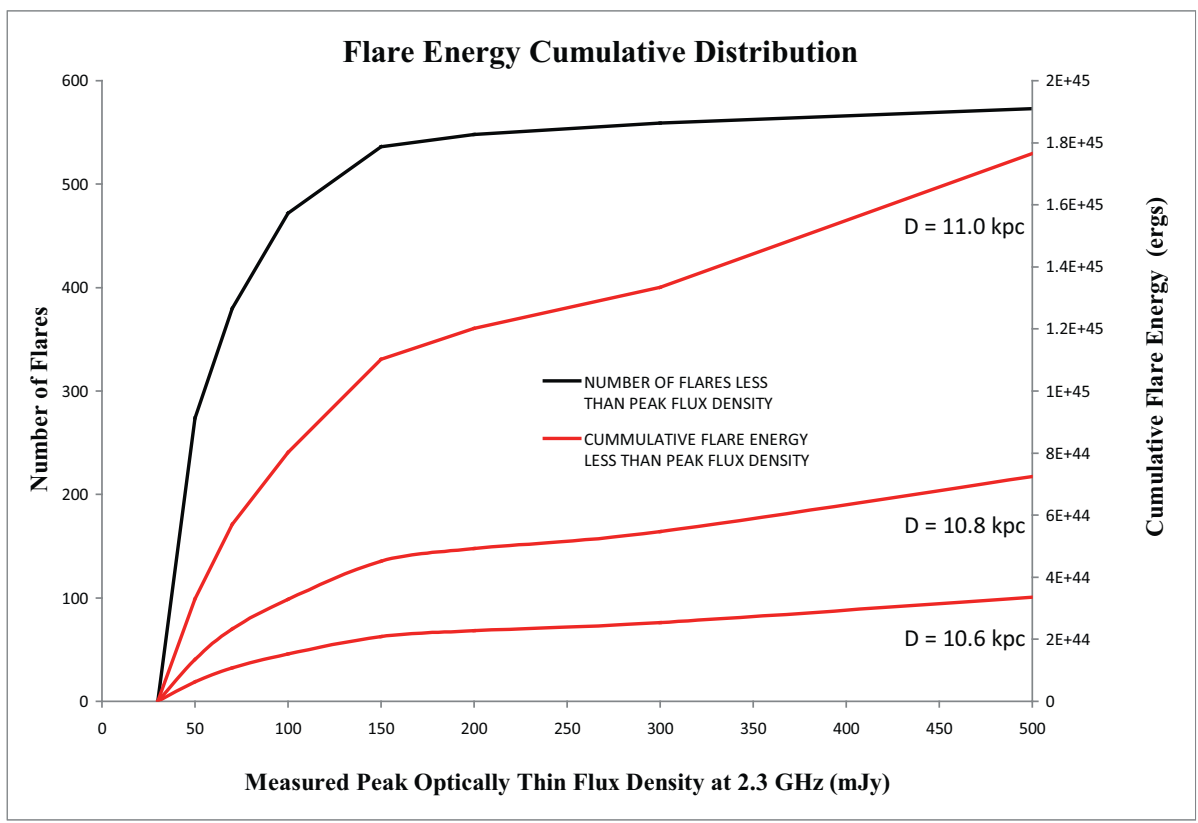}
\caption{The distributions of number and energy ejected in major
flares emitted from GRS 1915+105 from 1996 to the end of 2000 (see
text for details). The black (red) curve is the cumulative
distribution of flare number (energy) in the 1967 days,}
\end{figure}

\section{Scaling to a Supermassive Black Hole}
In this section, GRS~1915+105 is re-scaled to FR II quasar values of
$L_{\mathrm{bol}}$. A complete sample of quasars was created in
\citet{pun14} using the combined SDSS and FIRST databases. The radio
sensitivity was adequate to detect extended emission below the FR
I/FR II divide. Thus, a complete distribution of FR II quasars is
attained. Furthermore, the $\overline{Q}$ estimates are far more
accurate than other treatments of FIRST data in the literature.
Radio images were used to subtract the jet emission and core
emission on scales less than 20 kpc in order to more accurately
determine the optically thin lobe flux density which is the most
robust estimator for $\overline{Q}$ \citep{wil99}. The blue dots in
Figure 3 are the $\overline{Q}$-$L_{\mathrm{bol}}$ scatter plot from
the top left frame of Figure 3 in \citet{pun14} with the conversion
from the integrated optical/UV continuum luminosity to
$L_{\mathrm{bol}}$ given by Equation (3) of that paper. The
$L_{\mathrm{bol}}$ estimate is based on flux that is emitted $\sim
10^{6} - 10^{7}$ years after the preponderance of plasma that is
responsible for $\overline{Q}$ was ejected from the central engine,
based on lobe separation and estimates of lobe advance speeds
\citep{wil99}. Thus, the epochs are so displaced in time that there
need not be a causal connection. In order to compare GRS~1915+105 to
the AGN data, one needs to compare $\overline{Q}$ in GRS~1915+105 to
$L_{\mathrm{bol}}$ at epochs that are not likely to be causally
related, i.e. randomly selected from the historical distribution of
$L_{\mathrm{bol}}$.
\par This random data sampling is explored with Monte Carlo simulations. First, we create a distribution of quasar $L_{\mathrm{bol}}$ from the
SDSS/FIRST data scatter, $f_{\mathrm{quasar}}(L_{\mathrm{bol}})$.
Second, the $L_{\mathrm{bol}}$ distribution of GRS~1915+105 from
PR13 Figure 16 and $\overline{Q}$ from Equation (1) are combined to
create a random distribution of $\overline{Q}/L_{\mathrm{bol}}$,
$f_{\mathrm{GRS1915}}\left[\log\left(\frac{\overline{Q}}{L_{\mathrm{bol}}}\right)\right]$.
Then we create pairs of points representing the re-scaled
GRS~1915+105 based on re-scaling $L_{\mathrm{bol}}$ to the quasar
level. This is accomplished by randomly generating
$L_{\mathrm{bol}}$ from $f_{\mathrm{quasar}}(L_{\mathrm{bol}})$ then
creating $\overline{Q}$ randomly for each $L_{\mathrm{bol}}$ from
$f_{\mathrm{GRS1915}}\left[\log\left(\frac{\overline{Q}}{L_{\mathrm{bol}}}\right)\right]$.
The red dots in Figure 3 represent Monte Carlo simulations of the
re-scaled GRS~1915+105 with $L_{\mathrm{bol}}$ distributed similarly
to the SDSS/FIRST FR II sample. The explicit expressions for the
quasar luminosity distribution (the blue dots in Figure 3),
$f_{\mathrm{quasar}}(L_{\mathrm{bol}})$, is consistent with a
log-normal distribution, $Z$, with a mean logarithm (in units of
erg/s) of $45.96$ and a standard deviation, $0.29$,
\begin{equation}
f_{\mathrm{quasar}}[\log(L_{\mathrm{bol}})] = Z(\mu=45.96,
\sigma=0.29)\;.
\end{equation}
The distribution of $L_{\mathrm{bol}}$ for GRS 1915+105 is well
described by a log-normal distribution in Figure 16 of PR13
\begin{equation}
f_{\mathrm{GRS1915}}[\log(L_{\mathrm{bol}})] = Z(\mu=38.73,
\sigma=0.15);,\; D=11\, \mathrm{kpc}\;.
\end{equation}
From Equations (1) and (3), $\log(\overline{Q})= -1.71
+\mu[f_{\mathrm{GRS1915}}[\log(L_{\mathrm{bol}})]]$, thus
\begin{equation}
f_{\mathrm{GRS1915}}\left[\log\left(\frac{\overline{Q}}{L_{\mathrm{bol}}}\right)\right]
= Z(\mu = -1.71, \sigma=0.15)\;,\; D=11 \,\mathrm{kpc}\;.
\end{equation}
Similar expressions for $D < 11$ kpc follow from Table 2 and the
methods of PR13 (see Figure 2 also).

\par For D = 11 kpc the 2-D distribution of the re-scaled
GRS~1915+105 is clustered near the peak of the quasar distribution.
The smaller dispersion of the simulated data is expected because all
points are generated by the same central engine as opposed to a
variety of central BH masses and enveloping environments that are
responsible for the quasar generated scatter. At D = 10.7 kpc,
$\lesssim 1/2$ of the simulated data is consistent with FR II
quasars and at D= 10.6 kpc the distributions have become very
distinct. Systematic uncertainty is found by comparing the high
biased quasar $\overline{Q}$ estimates in Figure 3 from
\citet{wil99} with the low biased $\overline{Q}$ estimates from the
methods of \citet{pun05}, that are based on different assumptions.
This yields a systematic uncertainty in $\log{\overline{Q}}$ of
$0.38 \pm 0.08$, too small to affect the implications of the Monte
Carlo simulations. There is very little systematic error in
$L_{\mathrm{bol}}$ since it is derived from the integrated continuum
near the peak of the spectral energy distribution. The systematic
uncertainty in $\overline{Q}$ for GRS 1915+105 was discussed in
Section 2. The main systematic error in $L_{\mathrm{bol}}$ is in the
column density, $N_{H}$, to the source. From \citet{bel97,mun99} we
expect a systematic error less than a factor of 2 arising from the
uncertainty in $N_{H}$, too small to affect our conclusions.
\begin{figure}
\includegraphics[width=160 mm, angle= 0]{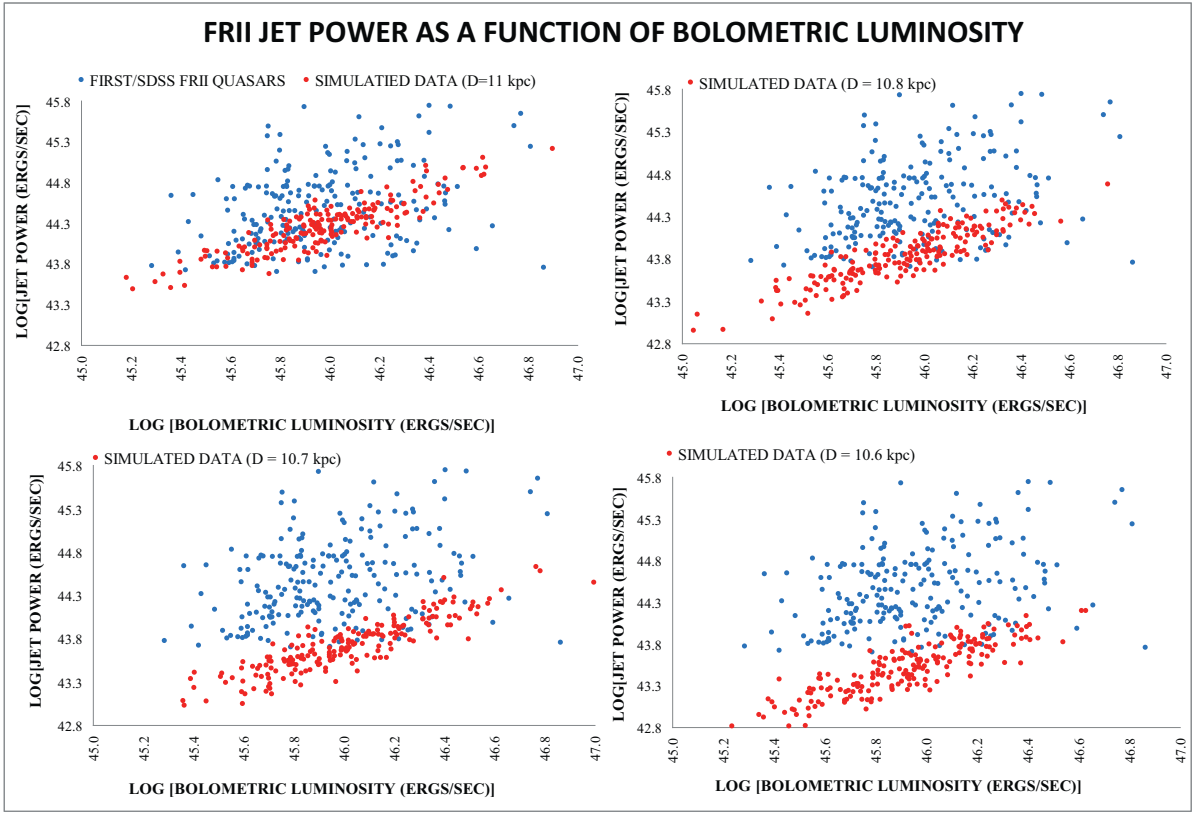}
\caption{Monte Carlo simulations are used to compare the 2-D
distribution in the $\overline{Q}$ - $L_{\mathrm{bol}}$ scatter
plane for a complete sample of SDSS/FIRST FR II quasars (blue) and
the re-scaled GRS~1915+105 (red)}
\end{figure}
\section{Results in the Context of Simulations of Black Hole Accretion}
We explore the main results from PR13 in the context of numerical
simulations
\begin{enumerate}
\item Strong flares are launched when $L_{\mathrm{bol}}$ is
at an elevated level (sometimes approaching the Eddington limit).
\item During the 1 to 7 hours of major flare ejections, the time
averaged power, $<Q>$, and time averaged intrinsic radiative
luminosity, $<L_{\mathrm{bol}}>$, are highly correlated.
\end{enumerate}
Our estimates of $L_{\mathrm{bol}}$ are based on models where the
main contribution is due to thermal Comptonisation of soft (cold
$\sim 0.2$ keV) photons by hot ($\sim$ 20-100 keV) electrons present
in a so-called corona. The important parameters to estimate
$L_{\mathrm{bol}}$ are $\mathrm{kT_{inj}}$, $\mathrm{kT_{e}}$,
$\tau$ and the Comptonised normalization (see Section 4.2.2 of
PR13). $<Q>/<L_{\mathrm{bol}}>$ for each D is computed from the
values in Tables 2 and 3 of PR13 for each flare. The solid line in
Figure 4 is $<Q>/<L_{\mathrm{bol}}>$ averaged over all the flares
for each D. The dashed lines represent the standard deviation,
$\pm\sigma$. The small dispersion argues strongly for an
approximately constant ratio $<Q>/<L_{\mathrm{bol}}>$ at each D. We
assume that this is the case and the errors in the individual $<Q>$
and $<L_{\mathrm{bol}}>$ are artifacts of imperfect data and our
estimation methods. Equivalently, the uncertainty in
$<Q>/<L_{\mathrm{bol}}>$ for each D is $\sigma$.

\begin{figure}
\includegraphics[width=125 mm, angle= 0]{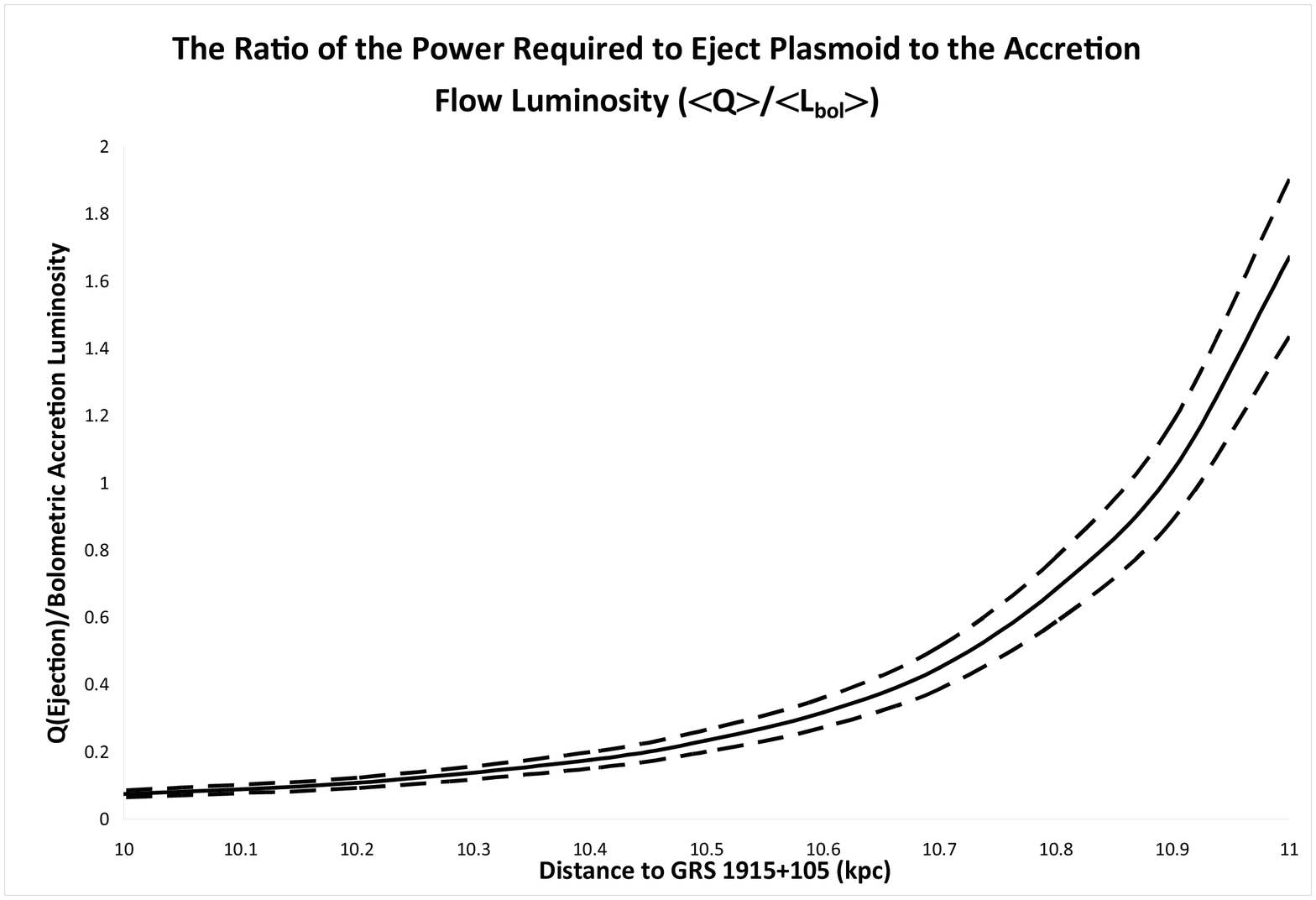}
\caption{$<Q>/<L_{\mathrm{bol}}>$ as a function of the assumed
distance, D, to GRS~1915+105. }
\end{figure}

There are two topologically distinct families of 3-D MHD simulations
of accreting gas near rotating BHs that produce relativistic
outflows. The limited flux simulations (LFS hereafter) evolve from
the accretion of weak dipolar loops of magnetic flux of the same
orientation from a finite torus of gas in the initial state
\citep{mck09,bec08,haw06,kro05}. In these simulations, only the
leading edge of the poloidal field accretes, so all the accreting
large scale magnetic flux is of one sign. In the LFS, it is the
accretion rate not the initial poloidal field strength that
regulates the long term, large scale magnetic field strength near
the black hole (and therefore the jet power) through ram pressure
in the inner disk (private communication McKinney (2011) and Tchekhovskoy in \citet{mar11})).
Physically, this situation might be considered a brief event of like sign large scale
flux that accretes to the BH and is maintained near the BH by the
dynamics of the accretion flow that is driven by a persistent MRI
(magneto-rotational instability). The physical appeal of LFS is that memory of the initial
state is erased and one finds a strong correlation between jet power and accretion rate in accord
with the findings of PR13 that were noted above.
\par The second type of simulation is
based on initial conditions that create MCAFs (magnetically choked
accretion) and MADs (magnetically arrested accretion)
\citep{mck12,tch11,tch12}. A new topology emerges, islands of large
scale magnetic flux perforate the disk, arresting the flow and
suppressing the MRI induced dissipation in these regions. Thus, one
would expect the luminosity to be suppressed compared to the LFS
\footnote{Simulations of MADs that were reported in \citet{pun11}
are generally subsonic and do not produce significant gas heating
from shocks.}. It was shown in PR 13 that $\approx 24$ hours before
an ejection $L_{\mathrm{bol}}$ is low and jetted emission is
minimal, thus it is not an MCAF/MAD state (which have maximal jet
efficiency). If an MCAF/MAD switches on to drive a major flare, it
would imply that the switch-on of an MCAF/MAD jet is preferentially
associated with an increase of X-ray luminosity to near the highest
historic (non-transient) levels. But this circumstance is
contradicted by the implications of our numerical simulations that
radiative efficiency should be suppressed in MCAF/MADs (suppression
was also noted in \citet{sik13}). Since MCAF/MADs simulations do not
appear to be representative of the high luminosity during major
ejections, in this article we concentrate on the LFS that can be
consistent with the dynamics of GRS1915+105.
\par In these simulations, the jet power, $Q$, is expressible in terms of
accretion rate onto the BH, $\dot{M} c^{2}$. Thus, one can compare
different simulations and one can compare to the observations if the
radiative efficiency of the accretion flow due to viscous
dissipation $\eta_{th}$, ($L_{\mathrm{bol}} \equiv \eta_{th}\dot{M}
c^{2}$) is known. In the numerical models, radiation effects are
simulated by ad hoc cooling functions that are based on the local
turbulent dissipation driven by MRI. In spite of a physically
incomplete methodology for treating the emissivity of the accreting
gas, the 3-D simulations of accretion disks in
\citet{pen10,nob09,nob11} have been used to estimate a value,
\begin{equation}
1.0\eta_{\mathrm{NT}}< \eta_{\mathrm{sim}}  \equiv \eta_{th} <
1.2\eta_{\mathrm{NT}}\;,
\end{equation}
where the range of results are expressed in units of
$\eta_{\mathrm{NT}}$, the value from \citep{nov73}. The disk
thickness in these simulations is in the range $0.05 < H/R < 0.2$.
We consider this range of $\eta_{\mathrm{sim}}$ from equation (5) as
a reasonable estimate for the $\eta_{th}$ of the LFS in Figure 5
since they have a compatible range of disk thickness, $0.15 < H/R <
0.2$ \citep{mck09,haw06}. Furthermore, it was demonstrated in the
simulations of \citet{sch12} that even if the detailed radiative
transfer results in the preponderance of disk luminosity being
created by Compton scattering in a disk corona instead of thermal
emission from the disk proper, $\eta_{\mathrm{sim}}$ is unchanged to
first order and is still consistent with Equation (5). This is an
important detail because X-ray spectra during the ejection of major
flares do not exist, so the relative contributions of coronal and
blackbody components are unknown. The simulations neglect radiation
pressure in the disk, so $\eta_{th}$ in Equation (5) might
not be accurate (but see \citet{szu96} who calculate $<10\%$ change
in $\eta_{th}$ for the relevant Eddington rates).
\par Figure 5 compares the simulated data to the relationship depicted in Figure 4.
The simulated data in the plot was presented and described elsewhere
(see \citet{pun09} for details). The only change is that the data is
normalized by $L_{\mathrm{bol}} = \eta_{\mathrm{sim}}\dot{M} c^{2}$
instead of $\dot{M} c^{2}$. The \citet{mck09} a/M=0.92 simulation
shows an event horizon jet as in \citet{blz77}. They note that $Q
\approx 0.01 \dot{M} c^{2}$ is similar to the 2-D solutions reported
in \citet{mck05}. Thus, the spin dependent $Q$ is given by equation
(3) of \citet{mck05}. This normalized jet power is plotted in green
in Figure 5. The other event horizon jet data (collectively referred
to as Event Horizon Jet (HK) in Figure 5) comes from
\citet{haw06,kro05} except for the raw data of \citet{bec07} for a/M
= 0.99 and a/M=0.998 \citep{pun09}. The ergospheric disk jet data is
from the simulations, KDE, KDH and KDJ from \citet{haw06,kro05}. The
nature and strength of these jets was described in detail in
\citet{pun11} and references therein. In order to remove the
artificial variation induced by computational grids covering
differing amounts of the ergospheric volume in different
simulations, a theoretical fit to the data, "ergospheric disk with
normalized inner boundary," (the red curve) was calculated in
\citet{pun09}. The fit was shown to exceed the KDE value because the
inner boundary of the computational grid is farther from the event
horizon in relative units (event horizon radius) than the other
simulations. Thus it does not sample the entire ergosphere (which is
critical for high BH spin in Boyer-Lindquist coordinates).
\begin{figure*}
\includegraphics[width= 185 mm, angle= 0]{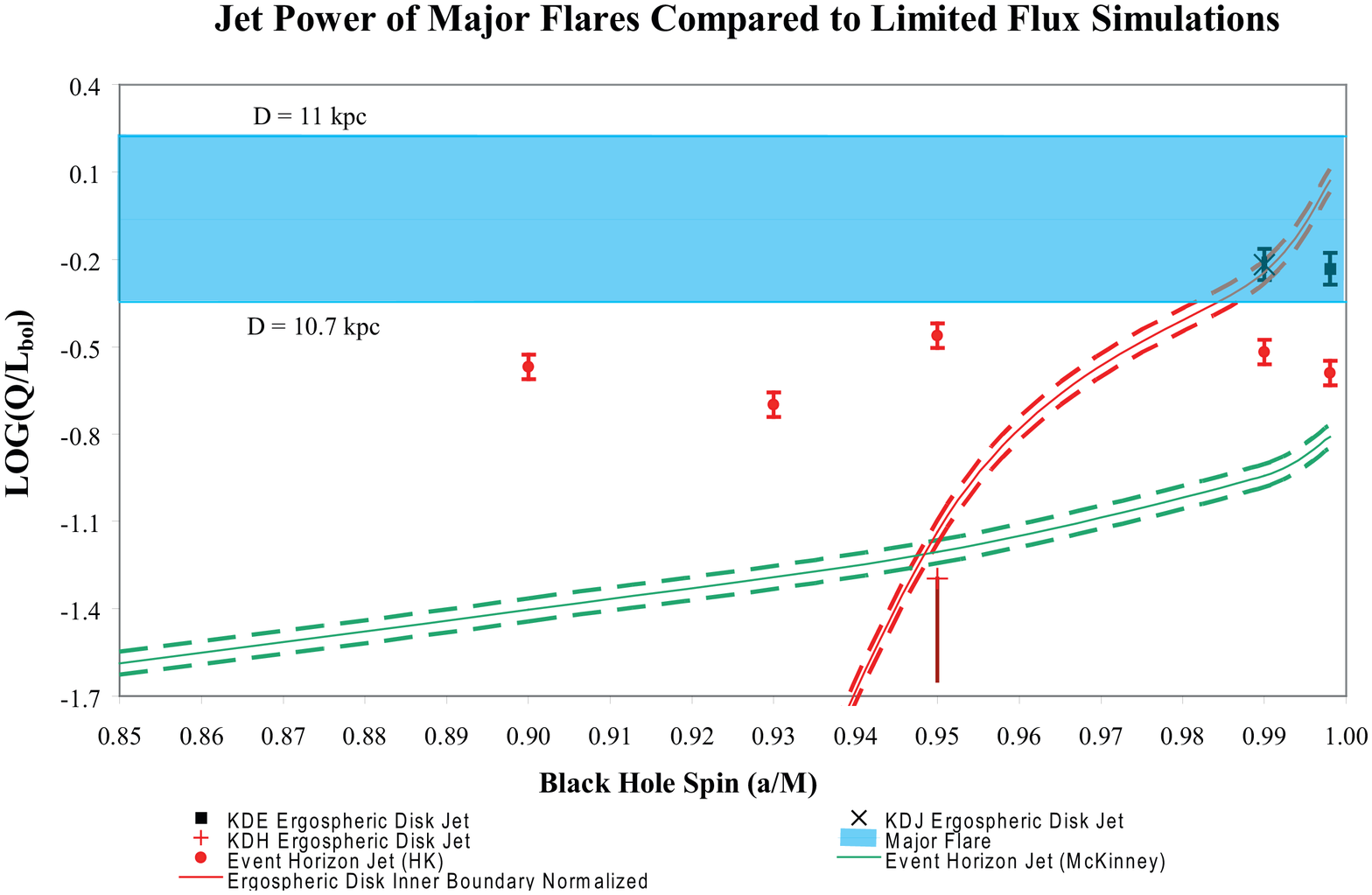}
\caption{The logarithm of "jet power/ $L_{\mathrm{bol}}$" in the
numerical simulations compared to $\log[<Q>/<L_{\mathrm{bol}}>]$ for
major flares in GRS~1915+105 based on Figure 4. The limited range of
10.7 kpc $<D<$ 11 kpc indicated by the blue band, is the range of
distance for which GRS 1915+105 can be considered a scaled down FR
II quasar. The error bars and the range of uncertainty on the curves
(the dashed curves) are based on the spread in Equation (5)}
\end{figure*}
The blue band in Figure 5 represents $<Q>/<L_{\mathrm{bol}}>$ from
Figure 4 corresponding to the viable range of $D$ that is compatible
with the scaling to FR II quasars in Figure 3, 10.7 kpc $< D < $
11.0 kpc (the dominant source of uncertainty is D). Agreement of the simulated data and observations is achieved for an
ergospheric disk driven jet with a black hole spin, $a/M = 0.99$ or $a/M=0.998$. The
theoretical fit in red continuously samples $a/M$ below 0.99 and
indicates agreement for $a/M > 0.984$. This result agrees with the
value of $a/M =0.99\pm 0.01$ that was estimated from the study of
X-ray spectra of GRS~1915+105 \citep{mcc06,blu09}.
\section{Conclusion} In this article, the long term and episodic behaviors of $L_{\mathrm{bol}}$
and the power of superluminal ejections in GRS~1915+105 were
compared and contrasted with FR II quasars under the assumption of
scale invariance of BH accretion systems. The results of Section 3
indicate that re-scaling $L_{\mathrm{bol}}$ of GRS 1915+105 to a
typical range of $L_{\mathrm{bol}}$ of FR II quasars yields a
consistent distribution of $\overline{Q}$ only if D $>10.7$ kpc -
otherwise the total energy emitted in the superluminal major
ejections is too small. Physically, this constraint on the distance
is equivalent kinematically to a Doppler factor $<0.48$ and a bulk
Lorentz factor $>3.5$ for the approaching plasmoids. Comparison of
the GRS~1915+105 data with the results of different 3-D numerical
simulations of the BH accretion indicate that GRS~1915+105 is
compatible with both numerical models and re-scaling to FR II
quasars if the numerical models contain an ergospheric disk jet and
if $a/M > 0.984$, in agreement with observational results obtained
by different groups and methods. If the analogy with FR II quasars
is robust and scale invariance is relevant to astrophysical BHs, the
results obtained for GRS 1915+105 may imply that typical radio loud
quasars also harbor rapidly spinning BHs.
\begin{acknowledgements}
 We would like to thank Igor Igumenshchev for many in-depth discussions
 on magnetically arrested accretion. We are also grateful to Matt Malkan for sharing his insight
into high Eddington rate accretion and the effects on the geometry
and efficiency of accretion disk models. JR acknowledges partial
funding from the European FP7 grant agreement number ITN 215212
``Black Hole Universe", and the hospitality of ESO (Garching,
Germany) where part of this work was done.
\end{acknowledgements}

\end{document}